\documentclass[fleqn,twoside]{article}
\usepackage[headings]{espcrc2}

\readRCS
$Id: espcrc2.tex,v 1.2 2004/02/24 11:22:11 spepping Exp $
\usepackage{graphicx}
\usepackage[figuresright]{rotating}

\newcommand{\AmS}{{\protect\the\textfont2
  A\kern-.1667em\lower.5ex\hbox{M}\kern-.125emS}}

\title{
 Infrared gluons, intrinsic transverse momentum and rising total cross-sections   \thanks{Presented at Hadron Structure '09, Tatranska Strba, September 2009. 
Thanks are due to the  Center for Theoretical Physics of the Massachusetts 
Institute of Technology, for hospitality while this work was being written. 
R.G. acknowledges support from Department of Science and Technology, India 
for financial support under Grant No. SR/S2/JCB-64/2007 (J.C. Bose fellowship)
This work has been partially supported by MEC (FPA2006-\-05294) and  Junta
de Andaluc\'\i a (FQM 101 and FQM 437).} 
}
    \author{A. Grau\address{Departamento de Fisica Teorica y del Cosmos, Universidad de Granada, Spain},
R.M. Godbole\address{Centre for High Energy Physics, Indian Institute of Science,Bangalore,
560012,India},
        G. Pancheri\address{INFN Frascati  National Laboratories, Frascati, I00444, Italy}
and        Y.N. Srivastava\address{Physics Department and INFN, University of Perugia,
            Perugia, Italy}}

\runtitle{ Infrared gluons, intrinsic transverse momentum and rising total cross-sections 
}
\runauthor{G. Pancheri}

\begin{document}
\def\sigmatot{$\sigma_{total}$}
\def\sigtot  {\mbox{$\sigma_{\rm tot}^{\rm pp/ p \bar p}$}}
\def\sigtotpp  {\mbox{$\sigma_{\rm tot}^{\rm pp}$}}
\def\rs{\mbox{$\sqrt{s}$}}
\def\pbarp{\mbox{$\rm  \bar{p}p$}}
\def\ptmin{\mbox{$p_{tmin}$}}
\def\pppbarp{pp-p{\bar p}}
\newcommand{\ra}{\rightarrow}
\newcommand{\ba}{\begin{array}}
\newcommand{\ea}{\end{array}}
\newcommand{\beqa}{\begin{eqnarray}}
\newcommand{\eeqa}{\end{eqnarray}}
\newcommand{\be}{\begin{equation}}
\newcommand{\ee}{\end{equation}}
\begin{abstract}
We discuss the infrared limit for soft gluon $k_t$-resummation and relate it to physical observables such as the intrinsic transverse momentum and the high energy limit of total cross-sections.
\vspace{1pc}
\end{abstract}

\maketitle
\section{Introduction \label{sec:intro}}
We present an analysis of the rise of total cross-sections achieved in 
our eikonal mini-jet model through an IR singular $\alpha_s$ and soft gluon 
resummation\cite{ourmodel}. We call it the Bloch-Nordsieck (BN) model because 
in it we include an infinite number of independently emitted very soft (IR) gluons.
Consider hadron-hadron scattering at a c.m. energy $\sqrt{s}$. In the 
eikonal representation,    the elastic and the total cross-sections  can be 
written as
\begin{eqnarray}
\sigma_{elastic}= \int d^2{\bf b} |
1-e^{
i\chi(b,s)
}
|^2 \\
\sigma_{total}=
2
\int d^2{\bf b} 
[
1-e^{
-{\cal I} m \chi(b,s)
}cos\Re e\chi(b,s)
]
\end{eqnarray}
The above two equations give for the total inelastic cross-section:
\begin{equation}
\sigma_{total\  inelastic}\equiv \sigma_{inel}=\int d^2{\bf  b} 
[
1-e^{
-2 {\cal I} m \chi(b,s)
}]
\label{sigmainel}
\end{equation} 
By relating the inelastic cross-section to the probability of all possible 
inelastic processes, considered independent of one another, one relates the 
average number of inelastic collisions to the imaginary part of the eikonal 
function, and the task is to adequately model this number.
Neglecting the real part of the eikonal (a good approximation for hadronic cross-sections at high energies), 
we obtain a simplified expression for the total cross-section:
\begin{equation}
\sigma_{total}=
2
\int d^2{\bf b} 
[
1-e^{
-{\bar n}(b,s)/2
}
]
\label{eq:sigtot}
\end{equation}
Quite some time ago, it was noticed\cite{halzen,levin} that perturbative QCD provides a 
simple mechanism for rising total cross-sections. The average number of collisions 
${\bar n}(b,s)$ increases because of the increasing number of low $x$ gluon-gluon collisions. 
These can be calculated perturbatively for all parton-parton processes with outgoing partons 
of $p_t>p_{tmin}$.

The cut-off $p_{tmin}$ performs a double r${\hat o}$le: (i) it avoids 
the Rutherford singularity
as $p_t\rightarrow 0$, as well as (ii) provides a scale above which 
perturbative  parton-parton cross-section estimates can be made using 
the asymptotic freedom (AF) expression  for the strong coupling constant 
$\alpha_s$. 

For the complete ${\bar n}(b,s)$, we need to add a non-perturbative part
\begin{equation}
 {\bar n}(b,s) = n_{NP} (b,s) + n_{hard} (b,s)
 \label{eq:nbs}
\end{equation}
where the  non perturbative (NP) term parametrizes the contribution of
 all those processes for which    initial partons scatter with 
$p_t<p_{tmin}$. We approximate the hard term,  which is
responsible for the high-energy rise and which we expect to dominate in the  
extremely high energy limit,
as
\begin{equation}
 n_{hard} (b,s) = A(b,s)\ \sigma _{jet} (s) 
  \label{eq:nhard}
 \end{equation}
 and calculate $\sigma _{jet} (s) $ using LO proton-proton cross-sections 
obtained from parton cross-sections and DGLAP \cite{DGLAP} evoluted Parton 
Density Functions of current use \cite{PDF}  at the scale $Q^2=p_t^2$.  
These cross-sections, when $p_{tmin} \approx 1\div 2 \ GeV$, have been  
called {\it mini-jets} to distinguish them from the high-$p_t$ jet 
cross-sections which 
are experimentally visible at high energies. These mini-jet cross-sections 
grow much too rapidly  with energy.
Imposition of unitarity in the eikonal impact parameter representation does indeed dampen in part this 
unacceptable growth. 

However, in order to properly reproduce 
the observed total cross-section high energy rise, from $\sqrt{s}\approx 10\div 20\ GeV$
to the Tevatron, Cosmic rays and extrapolations beyond, one needs 
to properly model the impact parameter dependence of partons in the hadrons. 
We shall discuss this in the coming sections. 

\section{Revisiting $k_t$ resummation
{\label{alphas}}}
\subsection{The infrared(IR) limit in  eikonal
 mini-jet models with soft gluon resummation}
 In our eikonal model for the 
total cross-section,  the rise is driven by perturbative QCD scattering
tempered by soft gluon resummation
down to
 IR region 
\cite{ourmodel}. To perform  resummation in this region, we used an expression 
for the effective quark-gluon interaction as $k_t^{gluon}\rightarrow 0$
 given by
\begin{equation}
\alpha_s(k^2_t)=\frac {p}{b_0\ln[1+p(\frac{k_t^2}{\Lambda^2})^p]}
\label{alphasint}
\end{equation}
where $b_0=(33-2N_f)/12 \pi$ is the  one-loop coefficient of the QCD beta 
function, and $\Lambda$ is the QCD scale. In the 
 $k_t^2>>\Lambda^2$ limit,
 the above expression reduces to the usual one loop asymptotic freedom 
expression for $\alpha_s$, whereas  
 $k_t^2<\Lambda^2$ limit allows 
integration into the IR region, 
provided $p<1$. In the  above expression, the constant 
$p$ in front of $k^2/\Lambda^2$ was included  to ensure that 
 $\alpha_s\ne 0 $ in the limit $p\rightarrow 0$. However, 
in \cite{froissartpaper}  we have subsequently found that 
in this model, one needs $p>1/2$ for analyticity of the scattering amplitude, 
so that the $p$ going to zero limit is 
never of interest,  and one could as well  use, for interpolation between the 
 IR and the UV region, a  
simpler  expression
\begin{eqnarray}
\alpha_s(k^2_t)=\frac {p}{b_0\ln[1+(\frac{k_t^2}{\Lambda^2})^p]}\\
\rightarrow \frac {1}{b_0\ln[\frac{k_t^2}{\Lambda^2}]}\ \ \ \ \ 
for \  \  \frac{k_t ^2}{\Lambda^2}\gg 1\\
\rightarrow \frac{p}{b_0}(\frac{k_t^2}{\Lambda^2})^p \ \ \ \ \ \  for 
\  \  \frac{k_t ^2}{\Lambda^2}\ll 1
\end{eqnarray}

\subsection{The impact parameter distribution}
 In our BN model,  
we have proposed that the impact parameter distribution of partons in hadrons 
be described by the Fourier transform of the soft gluon transverse momentum 
distribution, namely, for the average number of hard collisions, $n_{hard}(b,s)$, 
we have put
\begin{eqnarray}
\label{nbimp}
n_{hard}(b,s)=\sum_{i,j}\int{{dx_1}\over{x_1}}\times &&\nonumber \\
 \int{{dx_2}\over{x_2}} f_{/a}(x_1,p_t^2)f_{j/b}(x_2,p_t^2)
 \times&& \nonumber \\
   \int dz \int dp_t^2 
A_{BN}(b,M){{d\sigma}\over{dp_t^2 dz}}&&
\end{eqnarray}
where $f_{i/a}(x,p_t^2)$ are the parton densities
 in the colliding hadrons $a$ and $b$, evolved at the scale $p_t^2$, 
  $z={\hat s_{jet}}/(sx_1x_2)$, with $\sqrt{{\hat s_{jet}}}$ being 
the invariant mass  of the final parton-parton system emerging as two jets and 
 ${{d\sigma}\over{dp_t^2 dz}}$ is the
  differential cross-section for the process 
\be
\label{process}
parton + parton \to jet\ jet + X
\ee
The impact distribution function corresponding to the partonic collision is 
assumed to be given by 
\begin{equation}
\label{ourAB}
A_{BN}(b,M)=A_0 e^{-h(b,M)}
\end{equation}
with the normalization constant $A_0$  
\begin{equation}
A_0=\frac{1}{2\pi \int bdb \ e^{-h(b,M)}}
\end{equation}
and 
\begin{eqnarray}
\label{hdb}
h(b,M)=
{{2 c_F}\over{\pi}}
\int_0^M {{dk_\perp}\over{k_\perp}}\alpha_s({{k^2_\perp}
\over{\Lambda^2}})\times \nonumber \\
\ln{{M+\sqrt{M^2-k_\perp^2}}\over{M-\sqrt{M^2-k_\perp^2}}}
[1-J_0(k_\perp b)]
\end{eqnarray}
$M\equiv M(x_1,x_2,Q^2,s)$  is the maximum transverse momentum 
allowed  to single 
gluon emission by the kinematics of the process
\begin{equation}
parton(x_1)+parton(x_2) \rightarrow X (Q^2) + gluon (k)
\end{equation}
where X represents  a particle system with invariant mass $Q^2$, i.e.  two 
jets, for high energy parton scattering  at LO. The kinematics of the above process 
give \cite{greco}
\begin{equation}
M(x_1,x_2;Q^2,s)=\frac
{\sqrt{\hat s}} {2} (1-\frac{Q^2}{\sqrt{\hat s}})
\end{equation}
with ${\hat s}=4x_1x_2s$. We simplify the application of Eq. ~(\ref{nbimp}) by evaluating  
$A_{BN}(b,M)$ at a value $q_{max}$ which represents the average of M over all parton-parton 
processes, namely we shall use from now on, the factorized expression
\begin{equation}
n_{hard}(b,s)=A_{BN}(b,s)\sigma_{jet}(s,p_{tmin})
\label{eq:nhardav}
\end{equation}
where the $s$-dependence of $A_{BN}(b,s)$ obtains through 
 \begin{eqnarray}
<M(x_1,x_2;Q^2,s)> \equiv
q_{max}(s)=  {{\sqrt{s}} 
\over{2}}\times \nonumber \\
{{ \sum_{i,j}\int {{dx_1}\over{ x_1}}
f_{i/a}(x_1)\int {{dx_2}\over{x_2}}f_{j/b}(x_2)\sqrt{x_1x_2} \int dz (1 - z)}
\over{\sum_{i,j}\int {dx_1\over x_1}
f_{i/a}(x_1)\int {{dx_2}\over{x_2}}f_{j/b}(x_2) \int(dz)}}
\label{qmaxav}
\end{eqnarray}
with 
    the lower limit of integration in the variable $z$ given by 
 $z_{min}=4p_{tmin}^2/(sx_1x_2)$.
The scale parameter $q_{max}$ is a slowly varying function of $\sqrt{s}$ which depends on the PDF's 
used. In any phenomenological application, the PDF's used to evaluate 
$q_{max}$ will of course be of the same type as those used to evaluate  
$\sigma_{jet}$. Notice that  $q_{max}$ is of the order of $p_{tmin}$, since most of the parton-parton cross-section is peaked at $p_{tmin}$. In Eq.~(\ref{qmaxav}), we have dropped for simplicity the scale $p_t^2$ at which the densities are evaluated, but it is understood that all the densities are actually   DGLAP evoluted. It is through $q_{max}$  that $A_{BN}(b,s)$ acquires its 
energy dependence. This happens both from  $e^{-h(b,s)}$ as well as through 
the normalization constant $A_0$. We shall show how $A_0$ depends on the energy in a later section.

The  function $A_{BN}$ is obtained from the Fourier transform, 
${\cal F_{BN}}(K_\perp)$ of the 
transverse momentum distribution of the overall 
soft gluon radiation emitted (to LO) by quarks as the hadron 
breaks up because
of the collision. This distribution is obtained by summing soft 
gluons to all orders,
with a technique amply discussed in the literature \cite{pr76,ddt,pp}, namely
\begin{equation}
\label{BNPT}
{\cal F_{BN}}(K_\perp)={{1}\over{2 \pi}}\int b db J_0(b K_\perp) e^{-h(b,M)}
\end{equation}
As discussed in \cite{ourmodel}, 
we  use Eq.~(\ref{BNPT}) with  the soft gluon integration in  Eq.~(\ref{hdb}) 
extended well below the QCD scale $\Lambda$, where the asymptotic freedom 
expression for $\alpha_s$ is not valid. We enter this region, through  the 
 expression in  Eq.~(\ref{alphasint}).
In coordinate space, this $\alpha_s$ corresponds to a confining one-gluon exchange potential 
since it grows for large separation between quarks.

Using 
such an expression  allows us to push the $k_t$-integration in Eq.~(\ref{hdb}) 
down to zero values and hence access  the very  large distances 
which are relevant to  physical observables like the total cross-section or the intrinsic transverse momentum.

 We now   recall  how  Eq.~(\ref{hdb}) is commonly used. That is, one usually separates the IR region from the perturbative one as follows 
\begin{equation}
\label{h1}
h(b,E) = c_0(\mu,b,E)+ \Delta h(b,E),
\end{equation}
where
\begin{eqnarray}
\label{h2}
\Delta h(b,E) =\nonumber \\
 \frac{16}{3} \int_\mu^E {\alpha_s(k_t^2)\over{\pi}}[1- J_o(bk_t)]
 {{dk_t}\over{k_t}}
  \ln {
  {{2E}\over{k_t}}}.
\end{eqnarray}
 Since the integral in $\Delta h(b,E)$
now  extends down to a   scale $\mu \neq 0$, for $\mu> \Lambda_{QCD}$ one can use the asymptotic freedom expression for $\alpha_s(k_t^2)$.  Furthermore, having excluded the zero momentum region from the integration,  
 $J_o(bk_t)$ is  assumed  to oscillate to zero and    neglected.
The  integral of Eq.~(\ref{h2}) is now independent
of $b$ and  can be performed, giving
\begin{eqnarray}
\Delta h(b,E) =
\frac{32}{33-2N_f}\times \nonumber \\
\bigg\{ \ln ( {\frac{{2E}}
{\Lambda }} )\left[ {\ln ({\ln ( {\frac{E} {\Lambda }})}) -
 \ln
({\ln ( {\frac{\mu } {\Lambda }} )})} \right]
- \ln ( {\frac{E}{\mu }} ) \bigg\}. 
\end{eqnarray}
$\Lambda$ being the scale 
in the one-loop expression for $\alpha_s$.
In the range $1/E < b < 1/\Lambda$
the effective $h_{eff}(b,E)$
is obtained by setting $\mu = 1/b$ 
\cite{pp}. This choice of the scale   introduces a cut-off in impact 
parameter space which is stronger than any power, since
the radiation function, for $N_f=4$,  is now \cite{pp}
 \begin{equation}
e^{-h_{eff}(b,E)} =
 \big{[}
{{
\ln(1/b^2\Lambda^2)
}\over{
\ln(E^2/\Lambda^2)
}}
\big{]}^{(16/25)\ln(E^2/\Lambda^2)}
\label{PP}
\end{equation}
Under the assumption that there is no physical singularity in the range of 
integration
 $0\le k_t \le 1/b$, the remaining  $b$-dependent term, namely 
$exp[-c_0(\mu,b,E)]$,
is then dropped.  

By contrast,  it is the inclusion of the IR gluons, fortified by a singular but integrable $\alpha_s$, 
which shows up in our calculation as an energy independent smearing function phenomenologically 
called the intrinsic transverse momentum of partons.
The connection between $ c_0(\mu,b,E)$ and the intrinsic transverse momentum of 
partons is easily established formally if, in the region $bE\gg 1$ one 
makes the approximation \cite{nakamura,corsetti}
\begin{equation}
\label{intkdk}
h(b,E)\approx b^2\ A
\ee
with 
\begin{equation}
\label{A}
 A= {{c_F}\over{4\pi}}
\int  dk^2 \alpha_s({{k^2}\over{\Lambda^2}}) \ln
{{4 E^2}\over{k^2}}
\end{equation}
We obtain a function $h(b,E)$ which, through  Eq.~(\ref{ourAB}), 
gives 
  a gaussian fall-off as in models where 
  $A(b)$ is the Fourier transform of an intrinsic transverse momentum 
  distribution of partons, i.e.   $\exp(-k_\perp^2/4 A^2)$.
 We shall discuss this point further in the next section.
  \section{The intrinsic transverse momentum}
  The intrinsic transverse momentum is a phenomenological description of the 
very low-$p_t$ behaviour of hadrons, Drell-Yan pairs, $W-$mesons, jet-pairs, 
etc.,  produced in hadronic collisions. It was   discussed in 
\cite{greco,ddt,pp,nakamura} 
and  recently has been  studied phenomenologically in \cite{szcz}. It 
reflects the existence of a residual non-collinearity of quarks in the 
colliding hadrons, which cannot be estimated perturbatively through the 
Sudakov form factor.
Writing the contribution of the intrinsic transverse momentum as 
 $exp({-\frac{k_\perp^2}{<k_t^2>}}) $
and comparing with Eq.~(\ref{intkdk}) we have $p_{t-intrinsic}=\sqrt{<k_t^2>}=2\sqrt{A}$. In our model, in order to estimate a value for 
the intrinsic transverse momentum as a function of energy as done in 
\cite{corsetti}, the integration in the soft gluon momentum is pushed down 
to zero, using the singular but integrable expression for 
$\alpha_s$ presented in Sect.~\ref{alphas}. In Fig.~(\ref{fig:ptintrinsic}).
 the value of $p_{t-intrinsic}$ from \cite{corsetti} is plotted as a function 
of $M$, using $p=5/6$  following the argument in \cite{polyakov} about linearly rising Regge trajectories.
  \begin{figure}[htb]
\hspace{20pt}
\includegraphics[width=20pc,angle=90]{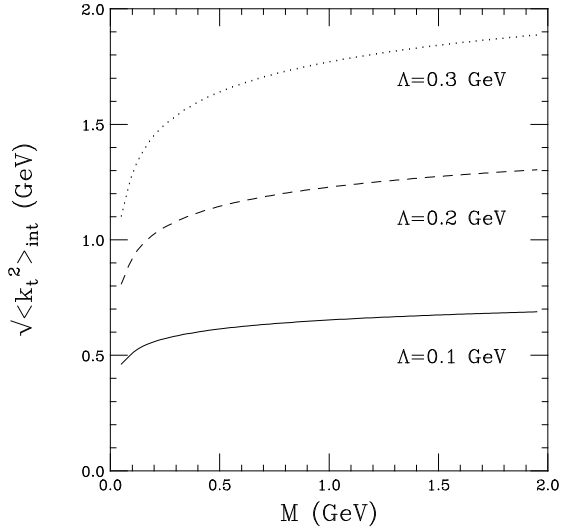}
\hspace{50pt}
\caption{ Intrinsic transverse momentum as a function of $M$ for different values of $\Lambda$,  $p=5/6$.}
\label{fig:ptintrinsic}
\end{figure}
In Fig.~(\ref{abfromff})  from \cite{corsetti} we reproduce the function $A_{BN}(b,s)$ 
for a range of values of the scale M, called here $q_{max}$. We also compare our proposed 
expression with the result for  the  Form Factor model,  in which the 
impact factor is independent of energy and obtained through  the convolution 
of the form factors of the colliding hadrons, namely
\be
\label{aff}
A_{FF}(b)={{1}\over{(2\pi)^2}}\int d^2{\vec q} e^{ib\cdot q} {\cal F}_1(q)
{\cal F}_2(q)
\ee
For protons,   the usual
 parametrization
\be
{\cal F}_{proton}(q)=\left(
 {{\nu^2}\over{q^2+\nu^2}}\right)^2\ \ \ \ \ \ \ \nu^2=0.71
GeV^2
\ee
leads to the following expression for 
the overlap function
\be
A_{FF}(b)={{\nu^2}\over{96 \pi}} \left(\nu b \right)^3 K_3(\nu b)
\ee
 \begin{figure}[htb]
\includegraphics[width=20pc,angle=90]{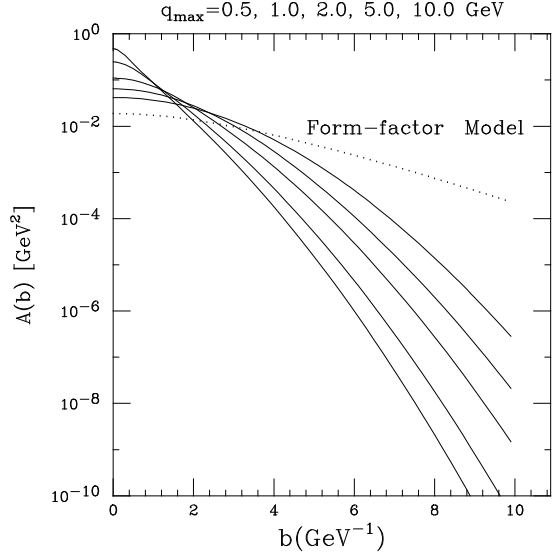}
\caption{ The A(b) distribution function from the Bloch-Nordsieck model 
for different $q_{max}$ values compared with the Form Factor model $A_{FF}(b)$, $p=5/6$.}
\label{abfromff}
\end{figure}

The expression for $\alpha_s$ in Eq.(\ref{alphasint}) allows to extend 
soft gluon resummation into the ultra-soft, zero momentum region. 
This can be done numerically, since    it is not possible to obtain an 
analytic expression for $h(b,M,\Lambda)$ valid in the full integration region. 
One can however divide  the integration region in various 
intervals, and make suitable approximations for the integrand.  
 Thus a   study of the $b$-region of interest, gives 
the result discussed in \cite{froissartpaper}, namely
\begin{equation}
b>{{1}\over{N_p\Lambda}}> {{1}\over{M}}
\end{equation}
\begin{eqnarray}
\label{halphas3} 
h(b,M) = 
 {{2c_F}\over{\pi}}\left[ {\bar b} {{b^2\Lambda^{2p}}\over{2}}
\int_0^{{1}\over{b}}{{dk}\over{k^{2p-1}}} \ln {{2M}\over{k}}+\right.  \nonumber \\
\left.  2 {\bar b} \Lambda^{2p}\int_{{1}\over{b}}^{N_p\Lambda} {{dk}\over{k^{2p+1}}}
\ln {{M}\over{k}}+{\bar b} \int_{N_p\Lambda}^M {{dk}\over{k}} 
{{\ln{{M}\over{k}}}\over{\ln {{k}\over{\Lambda}}}}\right]    \nonumber \\
 = {{2c_F}\over{\pi}} \Biggl [ {{{\bar b}}\over{8(1-p)}} (b^2\Lambda^2)^p
\left[ 2\ln(2Mb)+{{1}\over{1-p}}\right] +  \nonumber \\
+{{\bar b}\over{2p}}(b^2\Lambda^2)^p \left[2\ln(Mb)-{{1}\over{p}}\right]
+\nonumber \\
{{\bar b}\over{2pN_p^{2p}}}\left[-2\ln{{M}\over{\Lambda N_p}}+{{1}\over{p}}
\right] +   \nonumber \\
  {\bar b} \ln {{M}\over{\Lambda}}\left[\ln {{\ln{{M}\over{\Lambda}}}\over
{\ln{N_p}}}-1+{{\ln{N_p}}\over{\ln{{M}\over{\Lambda}}}} \right] \Biggr ]
\end{eqnarray}
where  $c_F=4/3$ for emission from quark legs, 
${\bar b}=12 \pi/(33-2N_f)$, and $N_p=(1/p)^{1/2p}>1$ for $p<1$.

Through this approximation, we see that our ansatz for 
$\alpha_s$ for $k^2/\Lambda^2 \ll 1$ leads to  the sharp cut-off in 
$e^{-h(b,M)}$ at large-$b$ values which we 
shall exploit to study the very large energy behaviour of the total
 cross-section in
our model, namely we obtain
\begin{equation}
A_{BN}(b,s)=A_0 e^{-h(b,M)}\approx e^{-(b{\bar \Lambda})^{2p}} \ \ \ \ \ b> 
\frac{1}{\Lambda}>\frac{1}{M}
\label{eq:elambda}
\end{equation}
Eq. ~(\ref{eq:elambda})  is similar to  Eq.~(\ref{intkdk}) for $p\approx 1$.
For the soft integral in $h(b,M)$ to be finite, however, $p< 1$  as one can see from the actual expression one obtains from ${\bar \Lambda}$ through Eq.~(\ref{halphas3}), namely
 \begin{equation}
 {\bar \Lambda(b,s)}=\Lambda 
 \{ 
 {{c_F{\bar b}}\over{4\pi(1-p)}}[
 \ln (2q_{max}(s)b) +{{1}\over{1-p}}]
 \}^{1/2p}.
 \end{equation}

 
 We plot in the next figures the values taken by $A_0$  as 
a function of energy for different cases. In Fig. ~(\ref{fig:a01})
we show how $A_0$ varies as function of
$q_{max}$ for proton-proton, for different values of the QCD scale $\Lambda$. 
Notice that , following our recent
phenomenological applications \cite{ourmodel,EPJC}, we have used $\Lambda=100\ MeV$ in the soft 
gluon integral, and $p\approx 0.75$.
  \begin{figure}[htb]
\includegraphics[width=15pc]{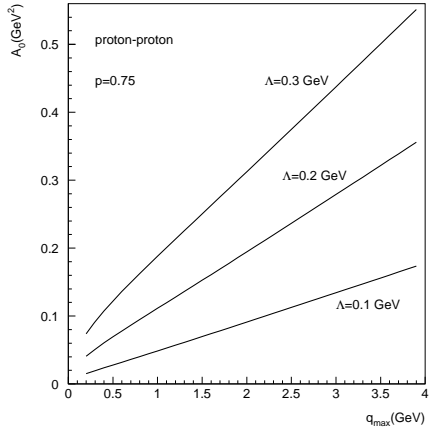}
\caption{ The normalization constant $A_0$ as a function of $q_{max}$ for different values of $\Lambda$.}
\label{fig:a01}
\end{figure}
 In Fig.~(\ref{fig:a02}) we show
$A_0$ as a function of $\sqrt{s}$ for the two cases of  proton-proton and 
$\gamma p$. The two cases differ because of  different values of the 
scale parameter $q_{max}$, which reflect different parton densities for 
photons and protons \cite{EPJC}.
 \begin{figure}[htb]
\includegraphics[width=15pc]{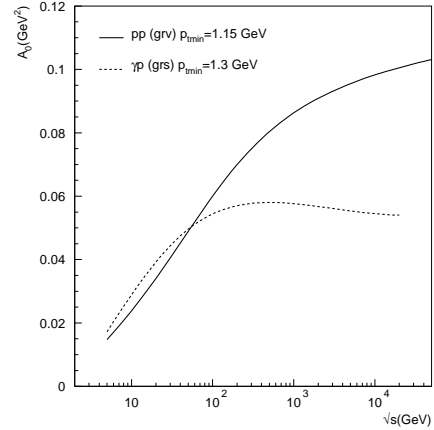}
\caption{The normalization constant $A_0$ as a function of $\sqrt{s}$ for the case of proton-proton and 
$\gamma p$ total cross-section calculation.  }
\label{fig:a02}
\end{figure}
We  also plot in Fig.~(\ref{fig:lambdabarppgp})
the energy dependence of ${\bar \Lambda}$, 
for proton-proton and for $\gamma p$,  
\begin{figure}[htb]
\includegraphics[width=15pc]{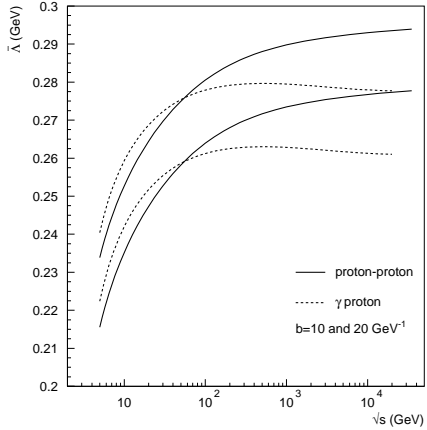}
\caption{${\bar \Lambda}$ as a function of energy, for photons and protons. }
\label{fig:lambdabarppgp}
\end{figure}
 obtained  through the 
parameter 
$q_{max}$ as mentioned before.

\section{The Froissart limit }
The sharp falling off  at very large $b$-values exhibited by our proposed impact parameter distribution can be exploited to discuss the large energy behaviour of the total hadronic cross-section in our model. Going to the very high energy limit in Eqs. (\ref{eq:sigtot},\ref{eq:nbs},\ref{eq:nhardav}), we can write
  \begin{equation}
  \sigma _T (s) \approx 2\pi \int_0^\infty  {db^2 } [1 - e^{ - n_{hard} (b,s)/2} ]
\end{equation}
Inserting  the asymptotic expression for $\sigma_{jet}$ at high
energies, which grows like  a power of $s$, 
and  $A_{BN}(b,s)$ from Eq.~(\ref{eq:elambda})
 we  obtain 
\begin{equation}
  n_{hard}  = 2C(s)e^{ - (b{\bar \Lambda})^{2p} }
\end{equation}
where  $2C(s) = A_0(s) \sigma _1 (s/s_0 )^\varepsilon  $. The resulting
expression for $\sigma_T$ 
\begin{equation}
  \sigma _T (s) \approx 2\pi \int_0^\infty  {db^2 } [1 - e^{ -
C(s)e^{ - (b{\bar \Lambda})^{2p} } } ]
\label{sigT}
\end{equation} 
leads to
\begin{equation}
{\bar \Lambda}^2\sigma_{T}(s)\approx (\frac {2\pi} {p})\int_0^{u_0} du u^{\frac{1-p} {p}}
=2\pi u_0^{1/p}
\end{equation}
with
\begin{equation}
u_0=\ln[\frac {C(s)}{\ln 2}]\approx \varepsilon \ln s
\end{equation}
To leading terms in $\ln s$, 
 we therefore derive the asymptotic energy dependence 
\begin{equation}
  \sigma _T  \to [\varepsilon \ln (s)]^{(1/p)}
 \label{froissart}
\end{equation}
 Since  $1/2<p<1$ \cite{froissart},  the above result shows that, with soft  gluon momenta  integrated into the IR region, $k_t<\Lambda$,   and a singular but integrable coupling to the quark current, our model leads to satisfaction of the Froissart-Martin bound \cite{froissart,martin}. This region, with  the   scale $\Lambda\simeq {\cal O}(\Lambda_{QCD})$ is of course  inaccessible to the  perturbative coupling for $\alpha_s$, but it plays a crucial role in many inclusive low $p_t$ processes.  One reason to neglect it could be that gluons with $|k_\perp|\ll \Lambda$ would see the hadron as a point-like object \cite{lipatov} and such emissions would have a small probability, because of colour screening.
 This argument is appealing, and similar to the one mentioned in Sect.~\ref{alphas}, but in our opinion, there is no compelling theoretical reason to assume that ultrasoft gluon emission in high energy reactions has low probablity. This argument     could be 
applied    to an isolated hadron, but not to 
  high energy hadronic scattering described through the scattering of partons, where
soft gluon emission is stimulated by QCD interactions.  It is through this interaction that we can expect 
the transition 
between hadrons and quarks to arise. 
A singularity in the infrared region 
would 
indeed provide a cut-off to separate quarks from hadrons and lead to such transition.
 This is the rationale behind going into the zero momentum region.

\begin{figure}[htb]
\includegraphics[width=20pc]{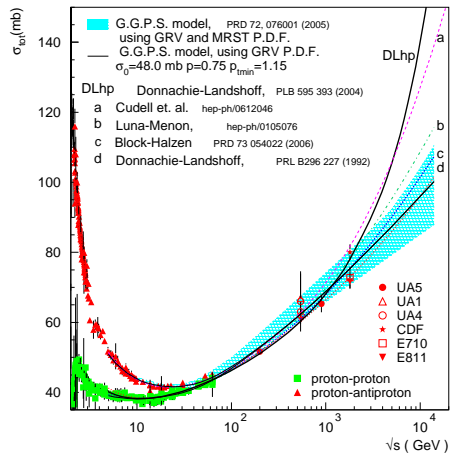}
\vspace{-20pt}
\caption{Data for the total proton-proton cross-section and comparison with the BN   and other  models \cite{pp_models}.   }
\label{fig:sigtot}
\end{figure}
We show in the last figure of this contribution a comparison between our model and the existing data for the total proton-proton cross-section. The band corresponds to a set of parameter values consistent with the discussion in the previous section, namely $p =0.75\div 0.8$, $p_{tmin}=1.15\ GeV$ and GRV  and MRST \cite{PDF} densities in the calculation of $\sigma_{jet}$ and $q_{max}$.

\section{Conclusions }
We have shown how a simple ansatz for the IR soft gluon spectrum allows to study the large impact parameter behaviour of some hadronic quantities, like the intrinsic transverse momentum and, most important, the total cross-section. Our ansatz relies on a power law behaviour for the coupling of very soft gluons to the quark current, which  makes it possible to integrate the soft gluon spectrum into the IR region. Our expression for the coupling is singular but integrable and interpolates between the AF and the IR region. 
\section*{Acknowledgments}
Thanks are due to the Organizers for this very interesting Conference and for hospitality. G.P is also grateful to  L. Lipatov for enlightening discussions.
 
\end{document}